\documentclass[aps,prl,twocolumn,superscriptaddress]{revtex4}
\usepackage{amsmath,epsfig,color,amssymb}
\usepackage{graphicx}
\usepackage{graphicx}
\usepackage{dcolumn}
\usepackage{bm}
\usepackage{graphicx}
\usepackage{amsmath, amsfonts, amssymb,mathrsfs}
\usepackage{pstricks}
\usepackage{amsxtra}
\usepackage{amsthm}
\usepackage{natbib}
\usepackage{csquotes}

\def\be{\begin{equation}}
\def\ee{\end{equation}}
\def\bea{\begin{eqnarray}}
\def\eea{\end{eqnarray}}
\newcommand{\ket}[1]{\mbox{$|#1\rangle$}}

\def\be{\begin{equation}}      
\def\ee{\end{equation}}
\def\beu{\begin{equation*}}   
\def\eeu{\end{equation*}}

\providecommand{\ket}[1]{\left|#1\right\rangle}

\providecommand{\del}{\partial}

\begin{document}
\title{Effective Field Theory for  Rydberg Polaritons}
\author{M. J.  Gullans}
\affiliation{Joint Quantum Institute and Joint Center for Quantum Information and Computer Science, National Institute of Standards and Technology and University of Maryland, College Park, Maryland 20742, USA}
%
\author{J. D. Thompson
}
\affiliation{Physics Department, Massachusetts Institute of Technology, Cambridge, Massachusetts 02138, USA}
\author{Y. Wang}
\affiliation{Joint Quantum Institute and Joint Center for Quantum Information and Computer Science, National Institute of Standards and Technology and University of Maryland, College Park, Maryland 20742, USA}
\author{Q.-Y. Liang}
\author{V. Vuleti{\' c}}
\affiliation{Physics Department, Massachusetts Institute of Technology, Cambridge, Massachusetts 02138, USA}
\author{M. D. Lukin}
\affiliation{Physics Department, Harvard University, Cambridge, Massachusetts 02138, USA}
\author{A. V. Gorshkov}
\affiliation{Joint Quantum Institute and Joint Center for Quantum Information and Computer Science, National Institute of Standards and Technology and University of Maryland, College Park, Maryland 20742, USA}

\begin{abstract}
 We develop an effective field theory (EFT) to describe the few- and many-body propagation of one dimensional Rydberg polaritons.  We show that the photonic transmission through the Rydberg medium can be found by mapping the propagation problem to a non-equilibrium quench, where the role of time and space are reversed.  We include effective range corrections in the EFT and show that they dominate the dynamics near scattering resonances in the presence of deep bound states.  Finally, we show how the long-range nature of the Rydberg-Rydberg interactions induces strong effective $N$-body interactions between Rydberg polaritons.    These results pave the way towards studying non-perturbative effects in quantum field theories using Rydberg polaritons.
\end{abstract}
\pacs{42.50.Nn, 32.80.Ee, 34.20.Cf, 42.50.Gy}

\maketitle

Photons can be made to strongly interact by dressing them with atomic Rydberg states under conditions of electromagnetic induced transparency (EIT) \cite{Pritchard10,Petrosyan11,Gorshkov11}.   Probing such Rydberg polaritons in the few-body limit, recent experiments were able to observe non-perturbative effects including the formation of bound states \cite{Firstenberg13}, single-photon blockade \cite{Peyronel12,Kuzmich12,Maxwell13} and transistors \cite{Hofferberth14,Tiarks14,Gorniaczyk15}, and  two-photon phase gates \cite{Tiarks16}.  
Theoretical work on quantum nonlinear optics with Rydberg polaritons has focused on two-body effects  or dilute systems \cite{Peyronel12,Petrosyan11,Gorshkov11,Firstenberg13,Buchler14,Maghrebi15,Hofferberth16,Otterbach13,Moos15,Grankin15}; however, these theoretical methods often fail in dense systems with more than two photons.

Effective field theory (EFT)  aims to describe low energy physics without resorting to a microscopic model at short distances or high energies \cite{EFTreviews}.  In few-body systems, it is a useful approach to describe particle scattering and bound states when the momentum $k$ involved is much less than the inverse range of the interactions \cite{EFTreviews,Braaten06} 
At the two-body level, the EFT depends only on the scattering length ${a}$. 
For scattering at momenta $k {a} \ll 1$, one can solve the EFT perturbatively \cite{Braaten97,Braaten06}.  
However, describing unitarity (${a} \to \pm \infty$) or bound states requires inclusion of 
 all orders in perturbation theory, which can be re-summed,
provided the EFT parameters are properly renormalized \cite{Adhikari95,Bedaque99b}.


In this Letter, we develop an EFT to describe the few- and many-body transmission of photons through a dispersive, one dimensional Rydberg polariton medium. 
We first consider the renormalized theory, which depends only on the local two-body scattering length, the effective mass, and the group velocity of the Rydberg polaritons.  By switching the role of time and space in the Lagrangian, we map the transmission problem to a non-equilibrium quench, which greatly simplifies the description of the dynamics.  
We then consider corrections to the EFT arising from the long range nature of the Rydberg interactions and the corrections to the massive dispersion.  We evaluate the so-called ``effective range corrections'' to the EFT and show that they dominate the dynamics near unitarity in the presence of deep bound states.  We then find the non-perturbative solution for the many-body Rydberg polariton problem at large momenta.   Integrating out this momentum scale leads to strong $N$-body interactions, which appear as contact forces in the EFT.

\begin{figure}[b]
\begin{center}
\includegraphics[width = .49 \textwidth]{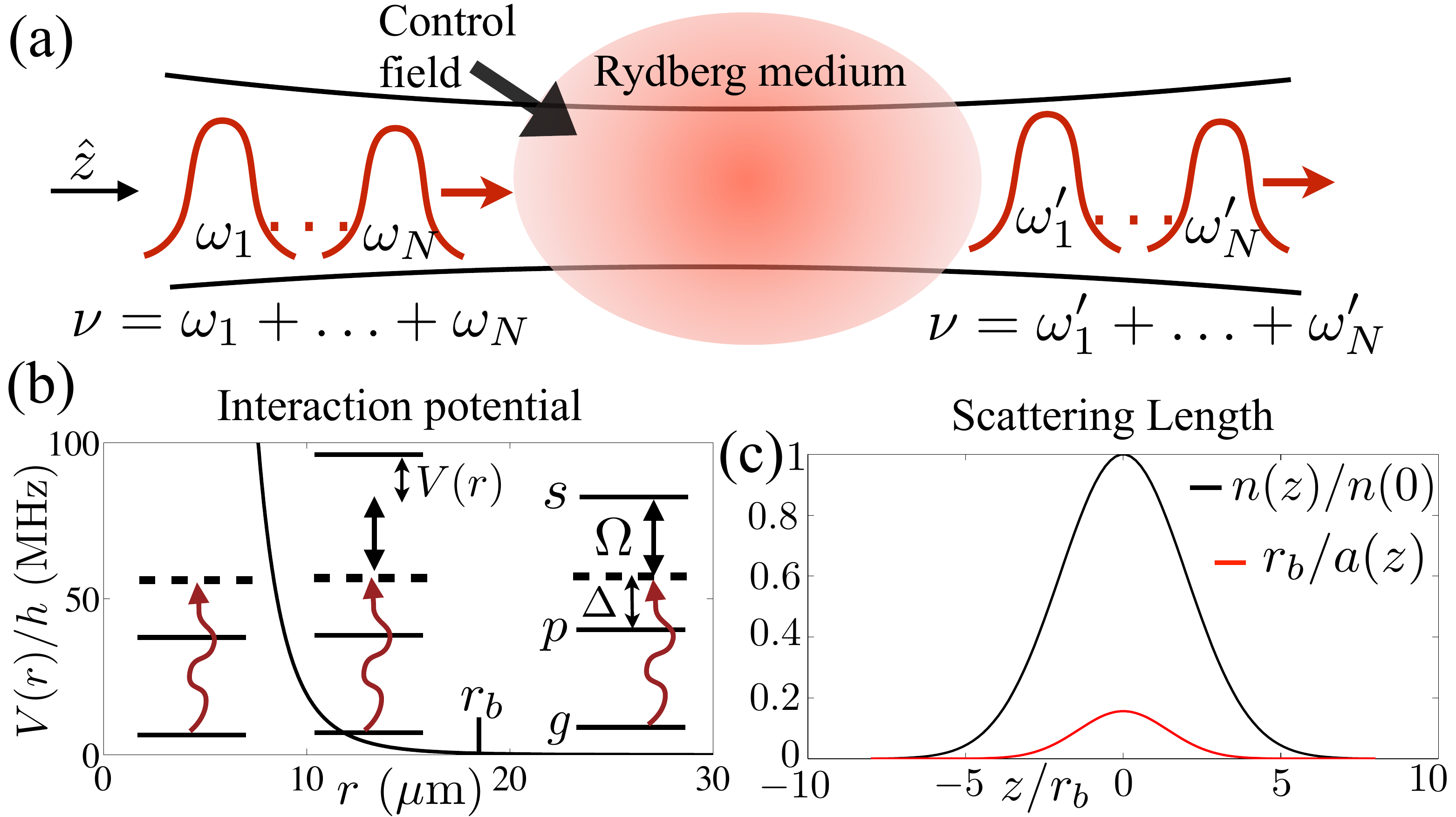}
\caption{(a) Rydberg polariton transmission experiment: an atomic cloud is probed with a few-photon beam, focused into a 1d channel, and a classical control field. Under dispersive conditions, the total energy  $\hbar \nu$ and number $N$ of probe photons are conserved. (b) Interaction potential $V(r)=C_6/r^6$ for two 100S$_{1/2}$ Rydberg states in {}$^{87}$Rb.  The range is given by the blockade radius $r_b$. (Inset) Level diagram of an  interacting atom for different $r$. 
 (c) Dimensionless density $n(z)/n(0)=\exp({-z^2/2 \sigma_{ax}^2})$ and inverse scattering length $r_b/{a}(z)$ for $\sigma_{ax}=36~\mu$m, $r_b=18~\mu$m,  ${\rm OD}=25$, $\Omega/2\pi = 5$~MHz, and $\Delta/2\pi = 20$~MHz.} 
\label{default}
\end{center}
\end{figure}

A schematic of a Rydberg polariton transmission experiment is shown in Fig.~1(a) \cite{Hofferberth16}. A spatially inhomogeneous atomic cloud is probed with a classical control field, with frequency $\omega_c$ and Rabi frequency $\Omega$, and a few-photon probe beam focused into a 1d channel.  The control and probe beams are configured for EIT on a two-photon resonance from the ground state $\ket{g}$ to a Rydberg state $\ket{s}$ via an intermediate state $\ket{p}$. We analyze the dispersive limit where the detuning of the control field $\Delta=\omega_{ps}-\omega_c$ is much greater than the $p$-state halfwidth $\gamma$; here $\omega_{ab}$ is the atomic transition frequency from $\ket{a}$ to $\ket{b}$. For large enough atomic density $n(z)$, the  probe photons  transform into Rydberg polaritons upon entering the medium  because the collective, single-photon Rabi frequency of the probe $g_c(z) =[6 \pi \gamma\, c^3 n(z)/\omega_{gp}^2]^{1/2}$ is much greater than  $\Omega$ \cite{Fleischhauer05}.  We use the dimensionless measure of the density given by the resonant optical depth ${\rm OD} =\int dz [g_c(z)]^2/2\gamma c$.

Consider two Rydberg atoms  interacting through the van der Waals potential $V(r) = C_6/r^6$. This interaction is strong enough that a single Rydberg excitation modifies the optical response over a region large compared to the optical wavelength (see Fig.~1(b) and inset).   
The size of this region is given by the blockade radius $r_b$, defined  by the condition that $V(r_b)$ is equal to the off-resonant EIT linewidth  $2 \Omega^2/|\Delta|$ \cite{Hofferberth16,footnote_rb}.  

To see how these effects lead to strong photon-photon interactions, one can use a gedanken experiment where one photon (polariton) is held at fixed position $z$, then any photon that passes by will pick up a nonlinear phase shift $\varphi(z) \approx   [g_c(z)]^2 r_b/c\Delta $;
here we assume $g_c(z)$ varies slowly over $r_b$.
 For atomic densities achievable with laser-cooled atoms ($n\gtrsim  10^{12}~$cm$^{-3}$), this nonlinear phase shift can be a sizable fraction of $\pi$ \cite{Firstenberg13}.  An alternative metric is the two-body scattering length ${a}$, which 
 was mapped out for Rydberg polaritons in a uniform medium in Ref.~\cite{Buchler14}.   For an inhomogeneous medium, we can similarly define a local scattering length ${a}(z)$.  For small $\varphi(z)$, these two metrics are closely related because ${a}(z) \approx (3/\pi)r_b/ [\varphi(z)]^2$.  We show in the supplemental material that ${a}(z)$ is well defined when the density varies slowly over $r_b$ \cite{supp}.   Figure 1(c) shows ${a}(z)$ calculated for a Gaussian density profile with parameters similar to recent experiments \cite{Firstenberg13}.

 In the absence of interactions, the propagation of Rydberg polaritons is captured by the local EIT dispersion relation \cite{Fleischhauer05,supp}
\be \label{eqn:eitq}
  q(\omega,z) = \frac{\omega}{c}\bigg(1+\frac{[g_c(z)]^2}{\Omega^2 - \omega(\Delta+\omega) } \bigg),
\ee
where $\omega=\omega_\ell -\omega_0$ is the detuning of the probe frequency $\omega_\ell$ from the two-photon resonance $\omega_0= \omega_{gs}-\omega_c$.  The electric field of the probe evolves as $E(\omega,z)=E(\omega,z_0) \exp\big[{ i \omega_0(z-z_0)/c+ i\int_{z_0}^{z}dz' q(\omega,z')}\big]$. For a sufficiently slowly varying density, we can define a local group velocity $v_g(z)= c/(1+[g_c(z)]^2/\Omega^2)$ and mass $m(z)=-{\hbar \Omega^2}/{2 \Delta [v_g(z)]^2} $ by solving Eq.~(\ref{eqn:eitq}) for $\omega$ and  expanding near $q=0$: $ \omega \approx v_g(z) q + \hbar q^2/2m(z)$ \cite{Fleischhauer05,footnote_mz}.  

For non-relativistic bosons in 1d, the only interaction term that is relevant under renormalization is the two-body contact interaction \cite{Adhikari95}.  As a result, the renormalized Lagrangian density for Rydberg polaritons is 
\be \label{eqn:L_z_old}
\mathcal{L} = \hat{\psi}^\dagger \Bigg[ i \hbar \del_t - {i\hbar v_g(z)}\del_z - \frac{\hbar^2 \del_z^2}{2 m(z)} \Bigg] \hat{\psi} - \frac{\hbar^2 \hat{\psi}^{\dagger2} \hat{\psi}^2}{m(z) {a}(z)} ,
\ee
where $[\hat{\psi}(t,z),\hat{\psi}^\dagger(t,z')]=\delta(z-z')$ and $\hat{\psi}$ is a single component field because there is only a single polariton branch near the two-photon resonance \cite{supp}.  Outside the medium, $\hat{\psi}$ is the quantum field for the probe photons, while inside it corresponds to the Rydberg polariton field. The scaling of the contact interaction as $1/{a}$  is the universal behavior for bosons in 1d, in contrast to higher dimensions where it scales as ${a}$ \cite{Olshanii98}.

Despite its relative simplicity compared to the microscopic model \cite{supp}, the theory is still difficult to solve because it has $z$-dependent parameters combined with second derivatives in $z$.  To overcome this  we  define a new EFT with time and space exchanged via the local transformation $(t,z) \to (z/ v_g(z),t v_g(z))$.     Similar transformations have been used to  study propagation of quantum light in nonlinear optical fibers \cite{Lai89a,Larre15}.  
For the steady state transmission with a uniform density, this transformation is equivalent to the rotated boundary conditions used in Ref.~\cite{Firstenberg13}.   
The resulting EFT is
\be \label{eqn:L_z}
\begin{split}
\mathcal{L}&=  \hat{\psi}^\dagger \Bigg[  i  \hbar v_g(z)\del_z - i \hbar \del_t  -  \frac{\hbar^2 \del_t^2 }{2m(z)[ v_g(z)]^2} \Bigg] \hat{\psi} \\
&- \frac{\hbar^2 \hat{\psi}^{\dagger2} \hat{\psi}^2  }{m(z) {a}(z)  v_g(z)} ,
\end{split}
\ee
where $[\hat{\psi}(z,t),\hat{\psi}^\dagger(z,t')]=\delta(t-t')$.   Up to higher order derivatives in $t$ (which can be neglected under renormalization), Eq.~(\ref{eqn:L_z}) is equivalent to Eq.~(\ref{eqn:L_z_old}); however, the second derivative is now in $t$ rather than $z$, which makes it easier to account for  the $z$-dependence of the  parameters. In particular, 
Eq.~(\ref{eqn:L_z}) gives rise to propagation equations akin to a time-dependent Schr{\" o}dinger equation
\begin{align} \label{eqn:Hz}
-i \hbar v_g(z) \del_z \hat{\psi}(z,t) &= \int dt' [ \mathcal{H}(z,t'),\hat{\psi}(z,t)] , 
\end{align}
where $\mathcal{H}(z,t)$ is given by the last three terms in Eq.~(\ref{eqn:L_z}).  In the dispersive regime, this propagation equation conserves the total photon number $N$, which simplifies the transmission problem.

\emph{Benchmarking the EFT.---}We now compare the predictions of the renormalized EFT for the two-photon transmission through a finite Rydberg medium with numerical simulations \cite{Peyronel12} of the  exact wavefunction propagation.   We decompose the  two-photon wavefunction at the exit of the medium $(z=L)$ as $\psi(L,t_1,t_2)= \sqrt{g^{(2)}(\tau)}e^{i \phi_2(\tau)+i 2 \phi_1}$, where $t_{1(2)}$ are the time coordinates of the two photons and $\tau=t_1-t_2$ is the relative time.  The probability density  $g^{(2)}(\tau)$ can be measured in two-photon coincidence  measurements of the output light for a weak coherent state input \cite{Peyronel12}, while the nonlinear phase $\phi_2(\tau)$ is defined relative to phase of the non-interacting medium with a single-photon phase shift $ \phi_1$ \cite{Firstenberg13}.

The results are shown in Fig.~2 for a representative set of parameters similar to Ref.~\cite{Firstenberg13}.
We take a steady-state probe input on two-photon resonance ($\omega=0$) with a Gaussian density profile $n(z) \propto \exp [{-(z-L/2)^2/2\sigma_{ax}^2}]$ with a cutoff at the entrance to ($z=0$) and exit from $(z=L)$ the medium.
   We  compare $g^{(2)}(\tau)$ and $\phi_2(\tau)$ found with three different methods: numerical simulations,  EFT with no free parameters, and EFT with a uniform density with $g_c$ a free parameter and medium length $L'$ chosen to match the time delay $\tau_d = \int_0^{L} dz[1/v_g(z)]$.  
  
  For an intermediate time window, we see that both EFT results capture many of the qualitative features of the simulations, but the inhomogeneous EFT captures more features and obtains better quantitative agreement.  We can understand the deviations at long and short times as follows. The long-time deviations arise because the EFT has a low momentum cutoff  associated with spatial variations in the density profile \cite{supp}.  For a Gaussian or uniform density profile, this scale is given by $1/L$, with the associated low-frequency cutoff $1/\tau_d$.  
  The short-time deviations arise from corrections to the EFT associated with: our use of a massive polariton dispersion, the swap of time and space, and the finite interaction range.  The first two effects contribute on timescales shorter than $\tau_m \approx \max(\Delta/\Omega^2,1/\Delta)$, while the effect of the finite interaction range appears on timescales less than  $r_b/v_g$.  
 For the parameters in Fig.~(2), $r_b/v_g, \tau_m \ll \tau_d$, which is consistent with the good agreement we find at intermediate times $r_b/v_g,\tau_m \lesssim \tau \lesssim \tau_d$.


\begin{figure}[t]
\begin{center}
\includegraphics[width = .49 \textwidth]{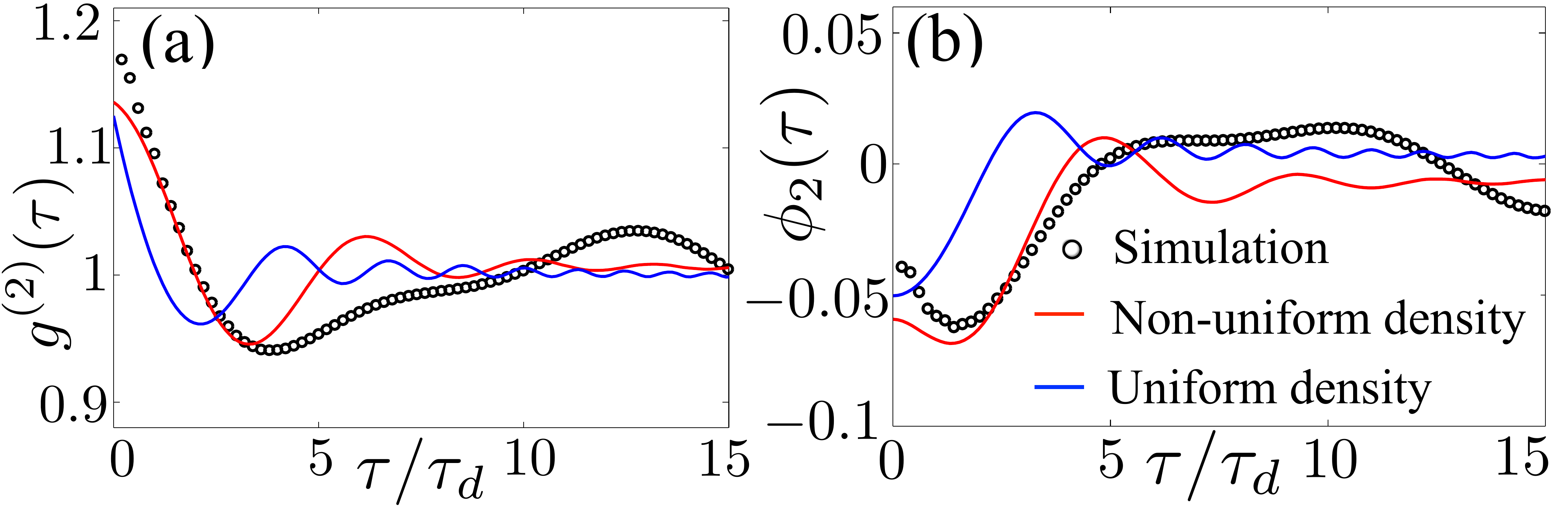}
\caption{  (a) Photon correlation function $g^{(2)}(\tau)$ and (b) phase $\phi_2(\tau)$ (in radians) of transmitted two-photon state calculated using EFT (solid lines) and numerical simulations (circles).  We took $\Omega/2\pi=5~$MHz, $\Delta/2\pi=20$~MHz, $r_b=10~\mu$m, ${\rm OD}=10$, $\sigma_{ax}=36~\mu$m, $L=4\, \sigma_{ax}$, and $L'=2.5\, \sigma_{ax}$.  To aid comparison we neglect decay from the $\ket{p}$ and $\ket{s}$ states.
} 
\label{default}
\end{center}
\end{figure}

In related work, we have shown that this renormalized EFT also gives good agreement with numerical simulations of the three-photon transmission \cite{Gullans16b}.  Yet, for increasing $N$, simulations of the full transmission become intractable and it
 is natural to ask: what are the leading corrections to the theory?  
In the framework of EFT, these corrections can be found systematically by evaluating higher order corrections in $k r_b$.   We show below that the terms in this expansion arise from two intertwined effects: (i) the finite range of the interactions and (ii) deviations of the dispersion from that of a massive particle.   
 
\emph{Effective range corrections.---}A standard approach to include finite range effects for massive particles is through the effective range expansion.  In this treatment,  higher order corrections to the scattering phase shift $\delta(k)$ are taken into account \cite{Braaten06}. For bosons in 1d, the expansion takes the form
\cite{Adhikari00}
 \be \label{eqn:delta}
 k \tan \delta(k) = \frac{1}{{a}} + \frac{r_0 k^2}{2} + \ldots,
 \ee
where $r_0$ is the so-called ``effective range'' parameter.
 These corrections can be included in the EFT by adding terms to the Lagrangian that contain higher derivatives in $\hat{\psi}$, e.g., (after switching time and space) 
\be \label{eqn:effrange}
\mathcal{L} \to \mathcal{L} + C_2 \, \hat{\psi}^\dagger  (\del_t \hat{\psi}^\dagger)(\del_t \hat{\psi}) \hat{\psi} ,
\ee
where $C_2 =\hbar^2 r_0/2m v_g^3$ is fixed by Eq.~(\ref{eqn:delta}) \cite{vanKolck99}.
Including these terms extends the validity of the EFT to higher polariton densities.  Most notably, this approach allows one to study unitarity in the presence of deep bound states, which occur when $\varphi \gg 1$.  
In this regime, we can solve for ${a}$ and $r_0$ analytically \cite{Gribakin93,Flambaum99}, and we find that the two-body contact vanishes near a scattering resonance, but $r_0 \approx 1.39 \sqrt{\varphi} \, r_b$  remains finite \cite{supp}.

 Scattering resonances associated with the appearance of additional two-body bound states can be achieved for Rydberg polaritons at sufficiently high atomic density \cite{Buchler14}.  
Current experiments, however, are limited to densities such that only a single two-body bound state is present.  In this case, we find $r_0 \approx (2/3) r_b^2/{a}$ and these corrections are suppressed.  We now show that the dominant corrections to the theory in this regime arise from effective 3-body interactions.

 \emph{$N$-body interactions.---}The strong long-range Rydberg interactions that result in blockade are also expected to induce large effective $N$-body interactions \cite{Buchler07,Weimer10}.
This is illustrated at the three-body level because,  when two polaritons are less than $r_b$ from a Rydberg atom, they do not interact with each other.  As a result, one expects a three-body force of the same magnitude and opposite sign as the two-body force.

More formally, effective $N$-body interactions emerge from integrating out virtual processes with high energy or large momenta.  In the case of Rydberg polaritons, this can be done in a surprisingly straightforward manner because the  theory dramatically simplifies at large momenta.  In particular,  the single-body propagator for the Rydberg polaritons projected onto the $s$-states $g_0^{ss}$ saturates to a constant (see supplemental material \cite{supp})
\be
\lim_{q \to \infty}\hbar g_0^{ss}(q,\nu)
= \hbar \chi(\nu)=\frac{(\Delta+\nu)}{(\Delta+\nu)\nu-\Omega^2},
\ee
for momentum $q\gg 1/v_g \tau_m$.
The physical origin of this can be seen in Eq.~(\ref{eqn:eitq}), where the local momentum $q(\omega,z)$ diverges at the Raman resonance conditions 
$
\Omega^2 - \omega (\Delta+\omega)=0.
$
  These Raman excitations have a frequency close to two-photon resonance and effectively infinite mass; therefore, near $\nu=0$, they dominate virtual processes with large internal momentum.    In the context of EFT, these effects can be included by adding two fictitious, infinitely massive particles to the theory associated with the Raman resonances \cite{supp}. 
Due to their high-energy, the Rydberg interactions can only excite these ``particles'' virtually.  
Integrating them out of the theory results in effective $N$-body interactions for the $\hat{\psi}$ field.  

The associated $N$-body interaction potential $V_\textrm{eff}^{N}$  can be found by accounting for all of the virtual processes where $N$ of these fictitious particles exchange momentum.  These contributions to the scattering amplitudes are represented by connected diagrams of the type shown in Fig.~3(a), where the particles cannot be broken into disjoint clusters.
Particles are connected by two-body interactions (curly lines), with the insertion of the $N$-body propagator in between (vertical lines).   
Figure~3(b) shows an example of a disconnected diagram, which is separable into two disjoint clusters $(12)$ and $(345)$ and does not contribute to $V_\textrm{eff}^N$.

\begin{figure}[t]
\begin{center}
\includegraphics[width = .49 \textwidth]{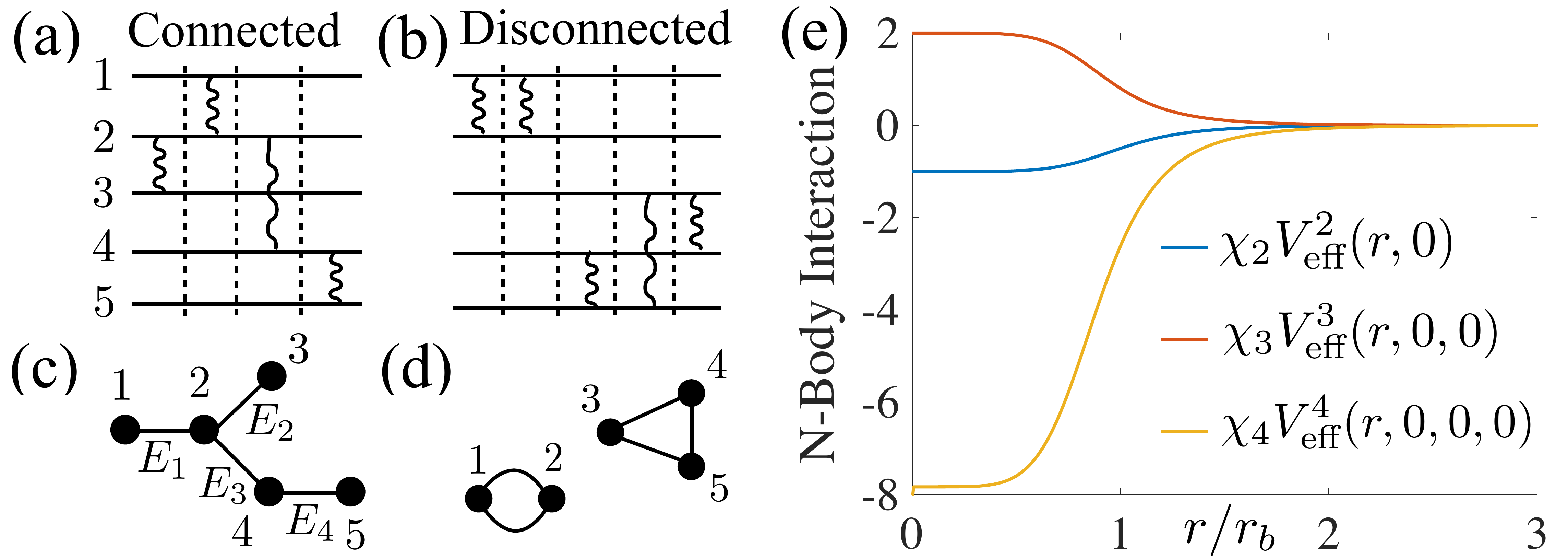}
\caption{Examples of a (a) connected and (b) disconnected scattering diagram for $N=5$.  The diagram in (a) contributes to $V_\textrm{eff}^N$, while (b) does not.  (c,d)  Graph representations of (a,b), respectively, neglecting the ordering of scattering events. The graph in (c) is a tree graph, which implies that (a) is a lowest order diagram for $V_\textrm{eff}^N$. (e)  Cut of the non-perturbative solution for $V_{\rm eff}^N$ in units of $\chi_N^{-1}$ up to $N=4$.} 
\label{default}
\end{center}
\end{figure}

Integral equations for the connected contributions to multi-particle scattering amplitudes were first formulated by Weinberg \cite{Weinberg64} and Rosenberg \cite{Rosenberg65}.   The full integral equations have only been solved for $N\le 4$ \cite{Kowalski91}; however,  the problem is simplified for the constant (i.e., momentum independent) propagator described above. The local nature of the propagator implies that the ordering of the scattering events is irrelevant.  In this limit, we can represent any scattering diagram by a graph of the type shown in Fig.~3(c,d), where the vertices represent particles and the edges indicate interaction pairs. Diagrams that map to a tree graph (e.g., Fig.~3(a,c)) give  the lowest order contribution to $V_\textrm{eff}^{N}$:
\be \label{eqn:pert}
V_\textrm{eff}^{N}(\bm{z};\nu) \approx (N-1)! [\chi_N(\nu)]^{N-2}  \sum_{T(N,E)} V_{E_1} ... V_{E_{N-1}},
\ee
where $E_k=(i_k,j_k)$ denotes a particle pair, $V_{E_k} = V(z_{i_k}-z_{j_k})$, and  the sum is over all labeled tree graphs $T(N,E)$ with $N$ vertices and $N-1$ edges $E=\{E_1,...\, ,E_{N-1} \}$.  Here the $N$-body propagator is $\hbar \, \chi_N(\nu)= \hbar^2\int d\omega \chi(\omega)\chi_{N-1}(\nu-\omega)\approx  ({\nu- N \Omega^2/\Delta})^{-1}$ (for $\Omega \ll \Delta$).  
If $r\ll r_b$, then $|V(r) \chi_N| >1$,  and the perturbative approach of Eq.~(\ref{eqn:pert}) is no longer valid.  We  derive the non-perturbative solution for $V_\textrm{eff}^N$ in the supplemental material \cite{supp}.  Figure~3(e) shows a cut of this solution up to $N=4$.  Consistent with the blockade effects described above, we see that $V_{\rm eff}^3$ has the opposite sign from $V_{\rm eff}^2$. More generally, we find $V_\textrm{eff}^N$ alternates with $N$ between attraction and repulsion  \cite{supp}.

During low-momenta processes $k r_0 \ll 1$, the polaritons hardly probe the blockaded region of the potential.  In this case, we can replace $V(r)$ with the renormalized interaction $U(r)=-(2 \hbar^2 /m {a}) \delta(r)$ and apply the perturbative result from Eq.~(\ref{eqn:pert}) to find the $N$-body interactions \cite{footnoteDelta}.
After switching time and space, the resulting EFT is governed by Eq.~(\ref{eqn:Hz}) with the Hamiltonian density
\begin{align} \label{eqn:Hn}
\mathcal{H}&=  \hat{\psi}^\dagger \bigg[-i  \hbar \del_t  - \frac{\hbar^2 \del_t^2 }{2m v_g^2} \Bigg] \hat{\psi} + \sum_N h_N \hat{\psi}^{\dagger N} \hat{\psi}^N,\\ \label{eqn:hn}
h_N& = \frac{(-1)^{N-1}}{N} \bigg(\frac{2\hbar^2}{m  {a}  v_g}\bigg)^{N-1}(N \chi_N)^{N-2},
\end{align}
where
 the two-body interaction $h_2$ is the same as in Eq.~(\ref{eqn:L_z}) and 
we used Cayley's tree formula $N^{N-2}$ for the number of labeled tree graphs with $N$ vertices in evaluating Eq.~(\ref{eqn:pert}) \cite{StanleyBook}.  
Using approximate expressions for ${a}$ near a scattering resonance \cite{supp}, we find the generic scaling $h_N\sim (r_b/{a})^{N-1}$.  

To determine the importance of the $N$-body interactions for non-perturbative effects in the EFT, $h_N$  
 should be compared with the effective range corrections at the momentum scale $k \sim 1/{a}$.   
 For large $\varphi$, $r_0 \sim r_b$ and, from Eq.~(\ref{eqn:effrange}), we see that the effective range corrections contribute at the same order as $h_3\sim r_b^2/a^2$. 
 On the other hand, for $\varphi \ll 1$, $r_0$ is suppressed by an additional power of $r_b/{a}$ and the effective range corrections scale as $h_4\sim r_b^3/{a}^3$.  Thus, for weak interactions, we find the surprising result that the 3-body force dominates the corrections to the theory for all momentum scales $k \lesssim 1/{a}$.

 The nature of these corrections has important implications for the propagation dynamics of 1d Rydberg polaritons.  The largest corrections to the theory will determine the deviations from the universal predictions for the shallow bound state clusters when ${a}>0$  \cite{McGuire64}, as well as deviations from the repulsive Lieb-Liniger model when ${a}<0$ \cite{Lieb63}.  In addition, all these corrections generically break the integrability of the EFT and, thus, determine the long time dynamics of the system \cite{Pric08,Imambekov10}.

\emph{Conclusion.---}
We  developed an EFT to describe the few- and many-body propagation of 1d Rydberg polaritons.   
The broad applicability of EFT to describe these systems opens a new perspective on the design of experiments aimed at probing non-perturbative effects in quantum field theories using Rydberg polaritons.  In particular,
Rydberg polariton experiments can involve complex geometries \cite{Sommer15} or more Rydberg levels \cite{Tiarks16},  dimensions \cite{Sevincli11}, and external control fields \cite{Maxwell14}.  The theoretical methods developed here can be naturally extended to these more complex configurations.   For example, 
using additional control fields or atomic levels to modify $\chi_N(\nu)$ would allow precise control over the range and strength of the $N$-body potentials. This could be used to realize exotic situations where, e.g., a single $M$-body force dominates over all $N$-body forces with $N \ne M$.  As another example, accounting for light diffraction introduces 3d effects, where EFT predicts the emergence of an Efimov effect in the vicinity of a scattering resonance \cite{Efimov71,Braaten06}.  
Further extending these theoretical methods to include dissipative interactions of the type demonstrated in Ref.~\cite{Peyronel12} may uncover new universality classes for few-body physics, as well as new  phases of non-equilibrium, strongly-correlated light and matter.

\emph{Note added.---}{During completion of this work, we became aware of related work on the three-body problem for Rydberg polaritons \cite{Buchler16}.}

\begin{acknowledgements}
\emph{Acknowledgements.---}We thank I.\ Carusotto, S.\ Diehl,  P.\ Julienne, B. Ruzic, O.\ Firstenberg, M.\ Maghrebi, and R.\ Qi  for helpful discussions.  This research was supported in part by the Kavli Institute for Theoretical Physics through the NSF under Grant No. NSF PHY11-25915, the NSF PFC at the JQI, the Harvard-MIT CUA, ARL CDQI, NSF QIS, AFOSR, and ARO.  
\end{acknowledgements}
\bibliography{RydEFT_Lett_resub_v1.bbl}

\begin{thebibliography}{50}
\bibitem[{foo({\natexlab{a}})}]{footnote_rb}
\bibinfo{note}{More precisely, we define $r_b$ via $V(r_b)=1/|\chi_2(0)| = 2
  \hbar \Omega^2 |\Delta|/|\Omega^2-\Delta^2|$ \cite{Buchler14}. For our
  parameters, this definition is nearly equivalent to the one in the main
  text.}
\bibitem[{foo({\natexlab{b}})}]{footnote_mz}
\bibinfo{note}{In the definition of $m(z)$ below Eq.~(1), we neglected a small
  correction of order $\Omega^2/[g_c(z)]^2$.}
\bibitem[{foo({\natexlab{c}})}]{footnoteDelta}
\bibinfo{note}{Here we are assuming the delta-function is regularized by a
  high-momentum cutoff $\Lambda$ satisfying $1/r_b \ll \Lambda\ll k$.}
\end{thebibliography}

\pagestyle{empty}
{ 
\begin{figure*}
\vspace{-1.8cm}
\hspace*{-2cm} 
\includegraphics[page=1]{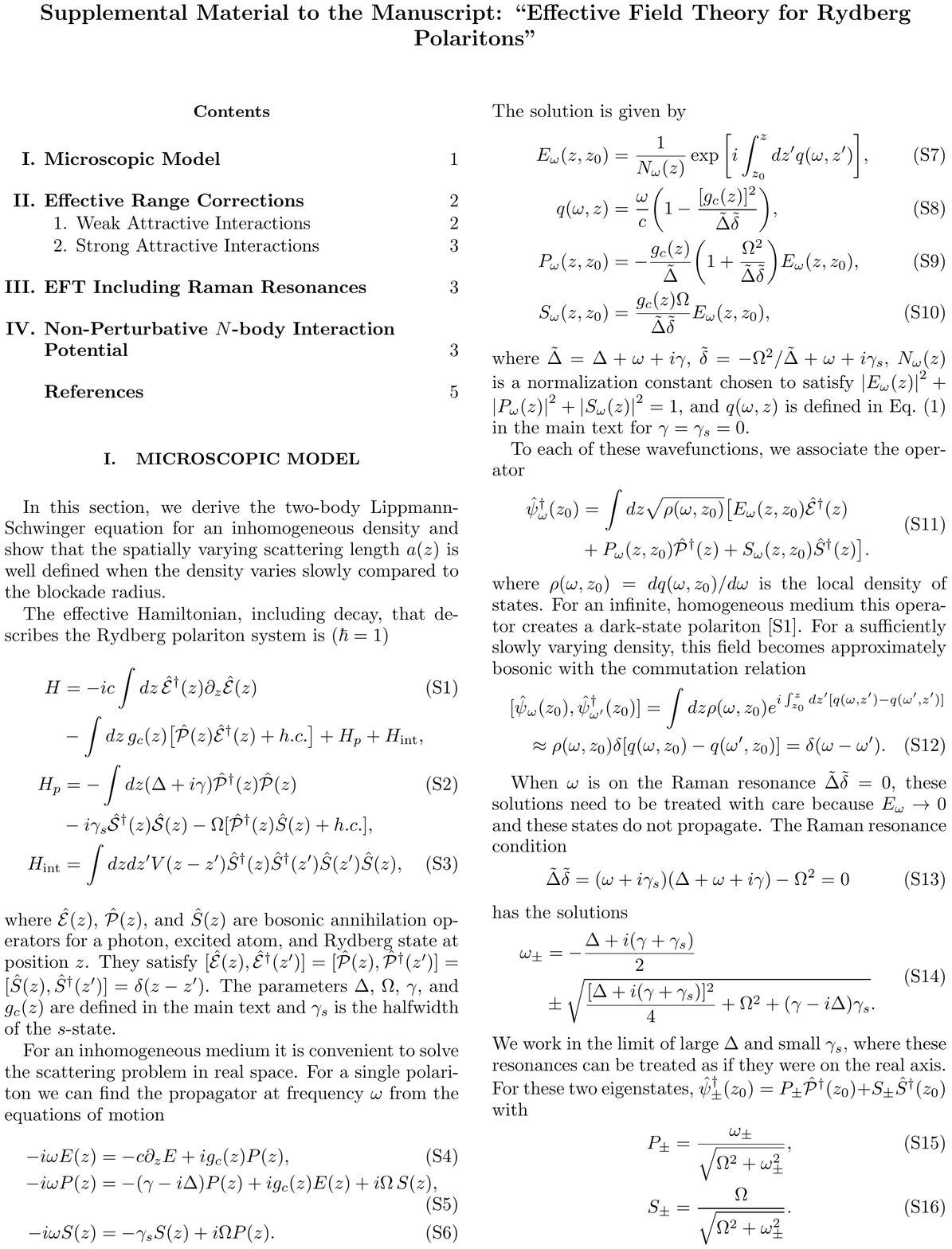}
\end{figure*}

\begin{figure*}
\vspace{-1.8cm}
\hspace*{-2cm} 
\includegraphics[page=2]{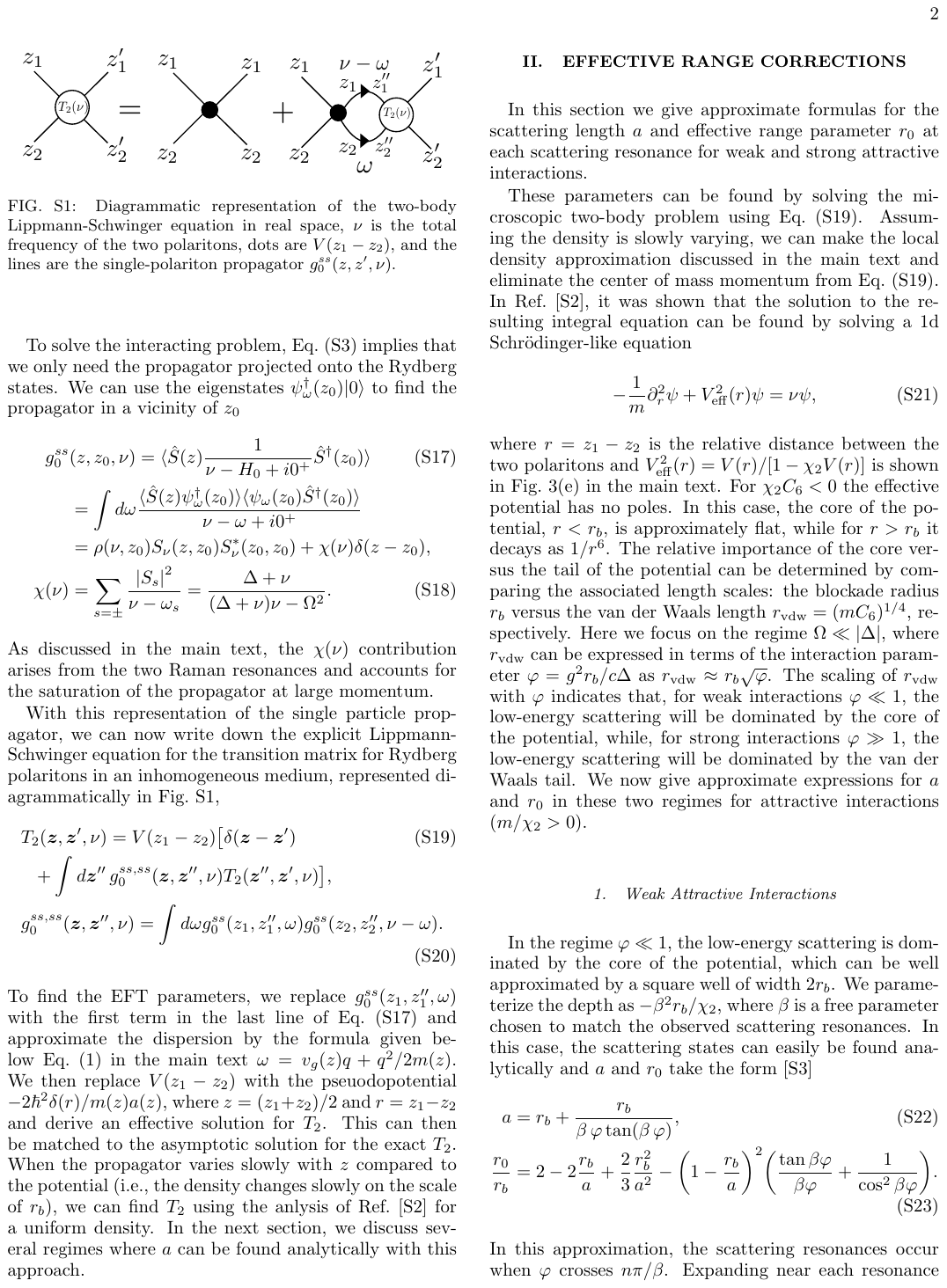}
\end{figure*}

\begin{figure*}
\vspace{-1.8cm}
\hspace*{-2cm} 
\includegraphics[page=3]{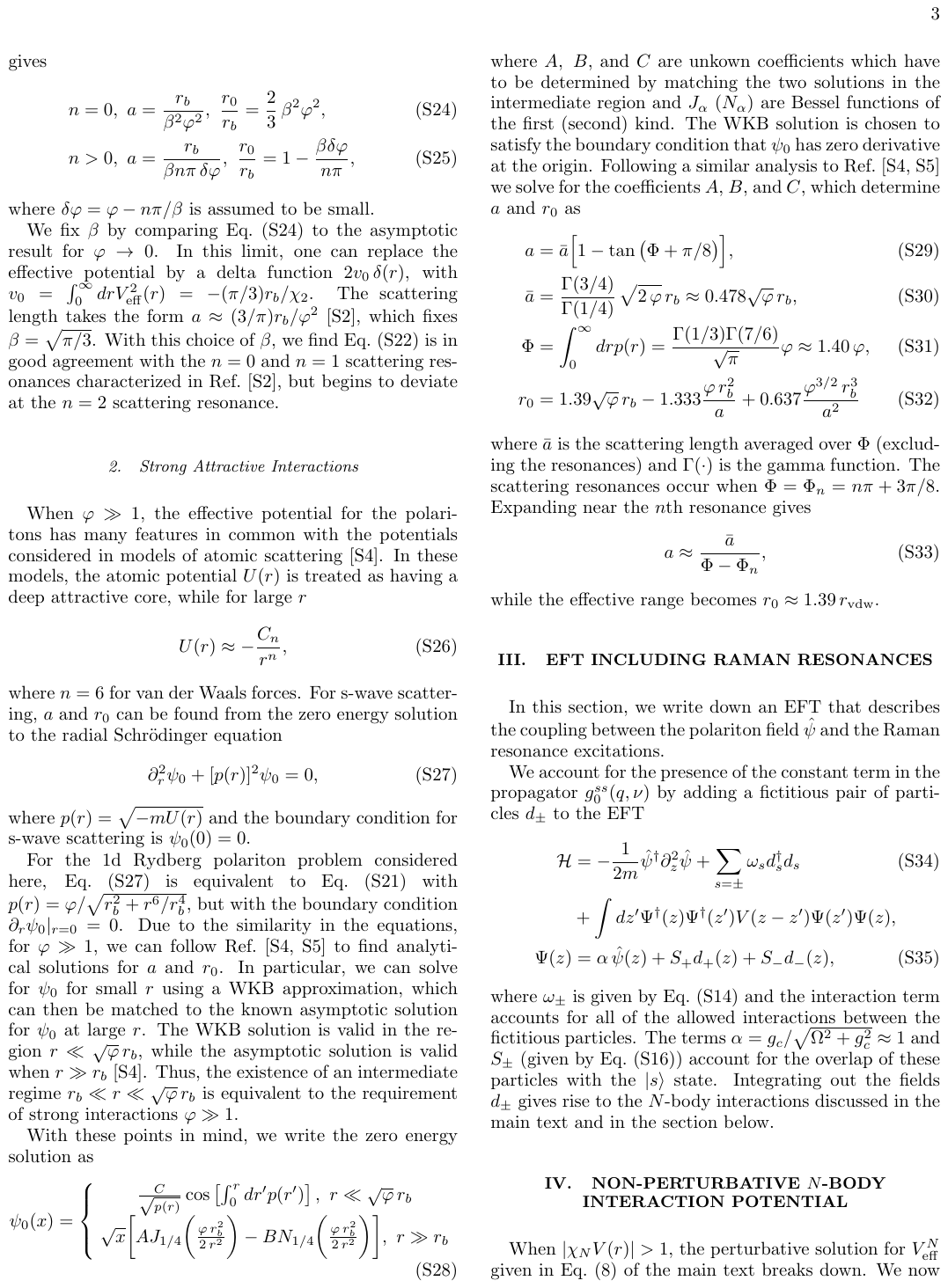}
\end{figure*}

\begin{figure*}
\vspace{-1.8cm}
\hspace*{-2cm} 
\includegraphics[page=4]{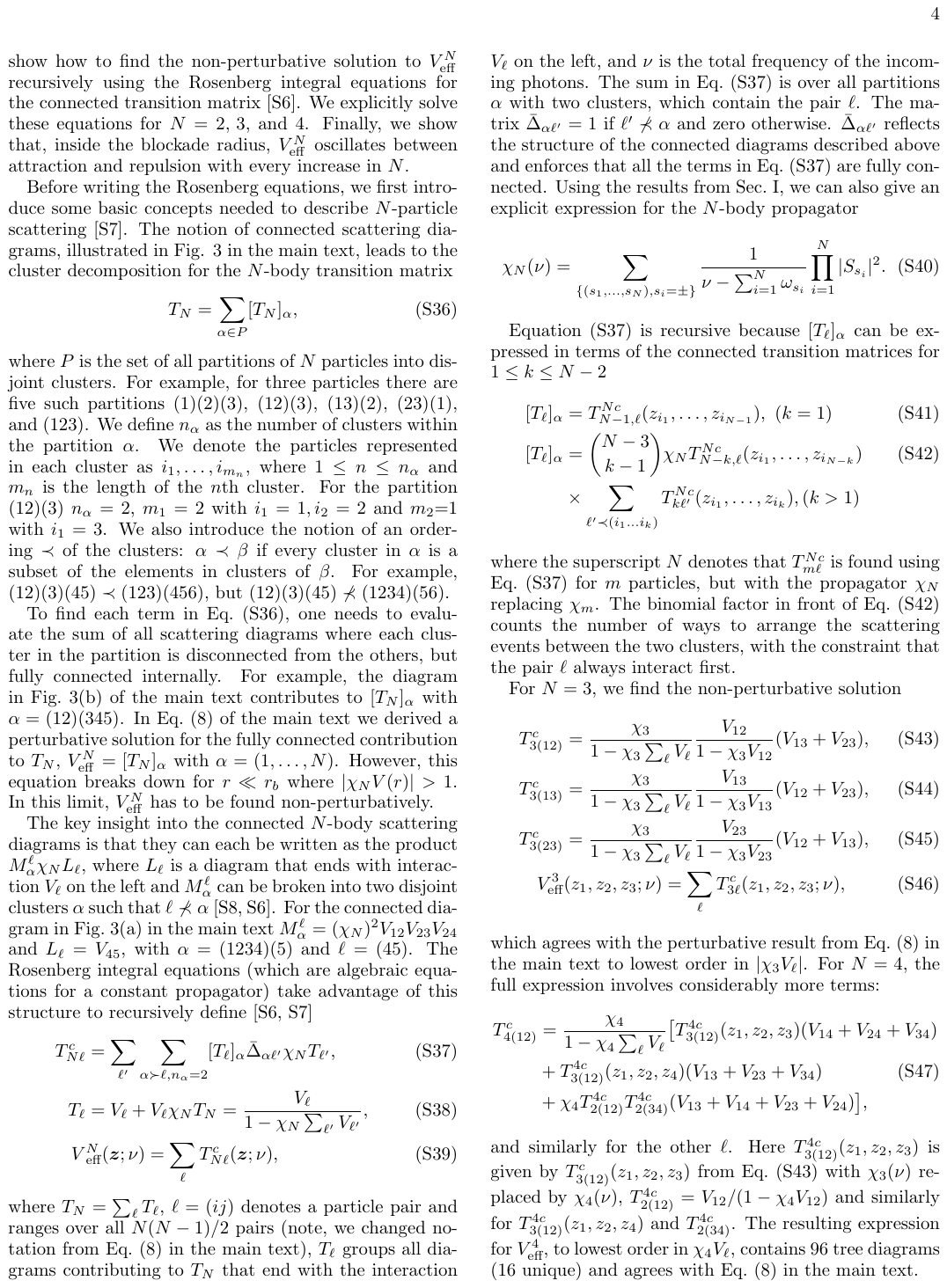}
\end{figure*}

\begin{figure*}
\vspace{-1.8cm}
\hspace*{-2cm} 
\includegraphics[page=5]{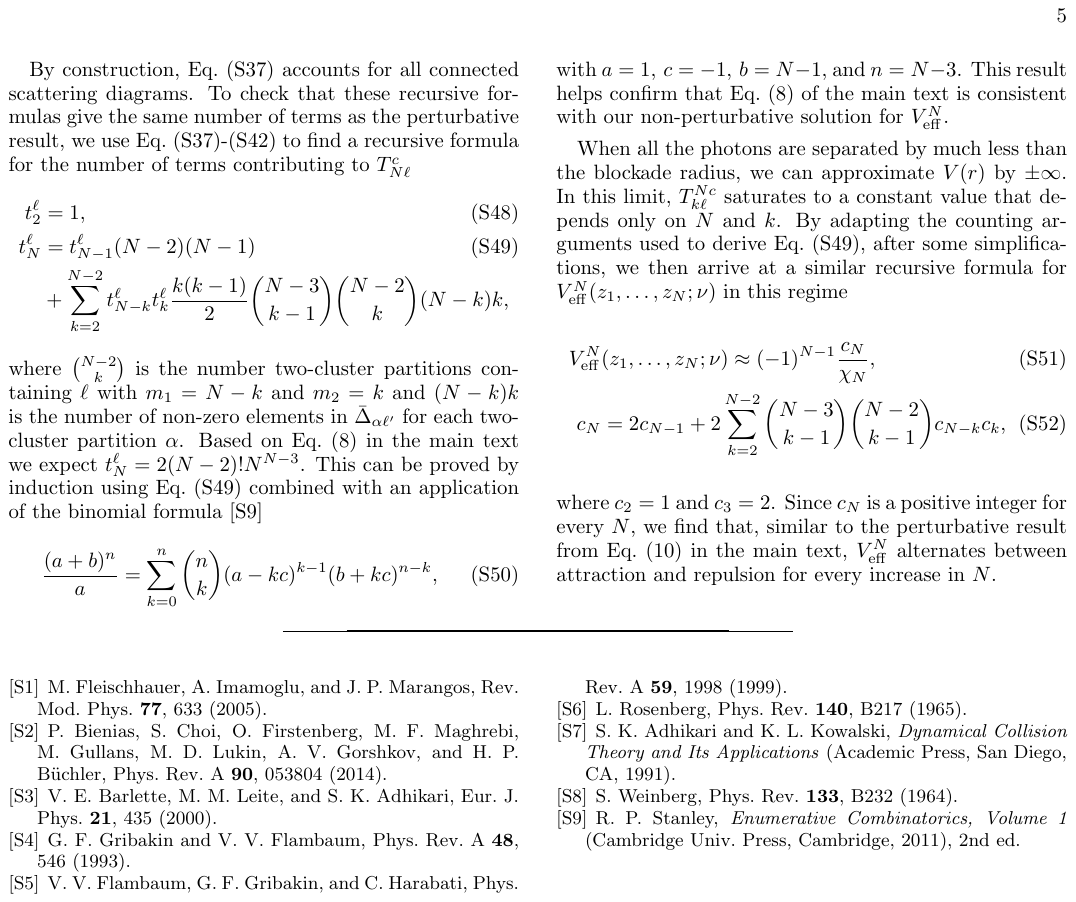}
\end{figure*}
}

\end{document}